\documentclass{Interspeech}
\usepackage{multirow}

\interspeechcameraready 

\usepackage{multirow}
\usepackage{colortbl}
\usepackage[table,xcdraw]{xcolor}
\usepackage{tabularx}
\usepackage{todonotes}
\usepackage{graphicx}
\usepackage{cite}

\newcommand{\bert}{\texttt{w2v-bert-2.0}\xspace}
\newcommand{\xls}{\texttt{wav2vec2-xls-r-2b}\xspace}

\title{TADA: Training-free Attribution and Out-of-Domain Detection of \\Audio Deepfakes}

\author[affiliation={1,2}]{Adriana}{Stan}
\author[affiliation={1}]{David}{Combei}
\author[affiliation={2}]{Dan}{Oneata}
\author[affiliation={2}]{Horia}{Cucu}

\affiliation{}{Technical University of Cluj-Napoca}{Romania}
\affiliation{}{University ``Politehnica'' Bucharest}{Romania}
\email{adriana.stan@com.utcluj.ro, david.combei@cs.utcluj.ro, \\dan.oneata@gmail.com, horia.cucu@upb.ro}


\begin{document}

\maketitle

\begin{abstract}
    Deepfake detection has gained significant attention across audio, text, and image modalities, with high accuracy in distinguishing real from fake. However, identifying the exact source—such as the system or model behind a deepfake—remains a less studied problem. In this paper, we take a significant step forward in audio deepfake model attribution or source tracing by proposing a training-free, green AI approach based entirely on k-Nearest Neighbors (kNN). Leveraging a pre-trained self-supervised learning (SSL) model, we show that grouping samples from the same generator is straightforward--we obtain an 0.93 F1-score across five deepfake datasets. The method also demonstrates strong out-of-domain (OOD) detection, effectively identifying samples from unseen models at an F1-score of 0.84. 
   
    We further analyse these results in a multi-dimensional approach and provide additional insights. All code and data protocols used in this work are available in our open repository: \url{https://github.com/adrianastan/tada/}.
    
\end{abstract}

\section{Introduction}

The rise of deepfake audio has raised concerns about authenticity and security, prompting progress in detection methods to distinguish real from synthetic audio~\cite{wang21fa_interspeech,kawa23b_interspeech,channing2024toward,truong2024temporal,pascu_is24, martindonas24_interspeech,pan24c_interspeech, chen24k_interspeech}. However, an equally critical task is \textit{model attribution} or \textit{source tracing}—identifying the specific generative model behind a deepfake. Unlike detection, which only confirms if audio is synthetic, attribution traces its origin, offering insights into the generator system and enabling targeted countermeasures. 
We define a \emph{generator} as a complete system comprising both the acoustic model and the vocoder. To emphasize that while different generators may share the same or somewhat similar architectures, it is the model weights that determine the exact source of the audio, for the evaluation section of this work we will predominantly use the term \textbf{checkpoint attribution}.

The task of model attribution or source tracing for spoofed or deepfake audio has only recently gained significant attention in the research community. Müller et al.~\cite{muller22b_interspeech} use either signal-based features (e.g., duration, jitter, pitch) or features derived from neural networks to perform clustering on both in-domain and out-of-domain attacks within the ASV19 dataset~\cite{asv19}. 
Zhu et al.~\cite{zhusource22} employ ResNet architectures for multi-task classification, focusing on waveform generator, conversion method, and speaker representation. 
All data used is in-domain, with a speaker-based split.
An important contribution to audio deepfake source tracing is the work of Klein et al.~\cite{klein2024source}, which combines various front-end architectures, such as ResNet, self-supervised learning, and processes them through AASIST~\cite{Jung2021AASIST} or the Whisper encoder~\cite{radford2022robustspeechrecognitionlargescale}. They propose multiple classification tasks, including identifying the acoustic model, vocoder, and input type (speech, text, or bonafide) to distinguish between TTS and VC models. However, they do not address the out-of-distribution (OOD) problem. The AASIST architecture is also applied in~\cite{xie24_interspeech} on top of SSL features for in- and out-of-domain classification with the ADD2023 dataset. The OOD detection utilizes a novel similarity metric.
 In terms of interpretable features and attributes, Phukan et al.~\cite{phukan2024investigatingprosodicsignaturesspeech} investigate the ability of various pre-trained models to capture prosodic information for deepfake attribution. 
In~\cite{yan2024rejectthresholdadaptationopenset}, Yan et al. propose a rejection threshold adaptation method for out-of-domain (OOD) detection. They convert closed-set classifiers into open-set classifiers and adaptively predict class-specific thresholds using Gaussian kernels. Xie et al.~\cite{xie2025neuralcodecsourcetracing} are the first to explore audio neural codec attribution, introducing a novel dataset of this kind. They propose methods for source tracing and OOD detection over this dataset using AASIST-derived architectures.

Building on previous work, we propose a novel method to identify the sources of generated audio across multiple datasets. Our approach is a training-free, green AI solution that requires no additional modules and leverages a simplified version of a pre-trained SSL model. 
The \textbf{contributions} of our work can be summarized as follows:
    (i) we introduce the first system for \textbf{model attribution across 5 multilingual audio deepfake datasets}, which also incorporates robust \textbf{out-of-domain detection} capabilities; 
    (ii) we propose a \textbf{highly accurate, ultra-light, computationally efficient, training-free approach} based on k-Nearest Neighbors (kNN), leveraging early features from a pre-trained self-supervised learning (SSL) model;
    (iii) we conduct an \textbf{extensive performance analysis}, demonstrating that the SSL features also capture attributes that pertain to the generative architectures and speaker identities;
    (iv) we highlight the challenges in \textbf{cross-dataset attribution}, revealing significant limitations in generalization across diverse models.

\vspace{-.2cm}
\section{Methodology}

\subsection{Audio Deepfake Datasets}

Table~\ref{tab:dataset} provides an overview of the audio datasets chosen for evaluating our work, with a brief description presented below. We emphasize that we selected those datasets where information about the generation method is available in the metadata or can be inferred from the dataset documentation and/or related publications. 

\textbf{ASVspoof 2019 (ASV19)}~\cite{asv19} serves as a benchmark for anti-spoofing research and has been extensively used for deepfake detection.   We use all 3 subsets: train, dev and eval.  
\noindent \textbf{ASVspoof 2021 (ASV21)}~\cite{asv21} expands on previous ASV19 and includes codec manipulate samples, as well samples coming from the Voice Conversion Challenges. As most of the data is already contained in ASV19, we only use the samples from the DF subsection which originate from the Voice Conversion Challenges 2018~\cite{lorenzotrueba2018voiceconversionchallenge2018} and 2020~\cite{vcc2020}. We also discard the data which was altered by codecs.
\noindent \textbf{ASVspoof 5 (ASV5)}~\cite{wang24_asvspoof, wang2025asvspoof5designcollection} is the latest dataset used in the ASVspoof Challenge. We use 
the complete train and test and select only the data which had no codec applied to it from the evaluation subset. 
\noindent \textbf{TIMIT-TTS (TIM)}~\cite{timittts} combines the TIMIT corpus with text-to-speech (TTS) technology to create a collection of synthetic speech samples. We select the CLEAN subset from it. 
\noindent \textbf{Multi-Language Audio Anti-Spoofing Dataset (MLAAD) v5}~\cite{muller2024mlaad} represents an essential resource as it also contains multispeaker, multilingual checkpoints, as well as some checkpoint overlap with other datasets (as we will show in the following sections). MLAAD does not include details on the vocoders and speakers. 

For all datasets we discard the bonafide samples. 
A total of 741k fake samples coming from 243 (presumably) different text-to-speech synthesis (TTS) or voice cloning (VC) model checkpoints are used. 

\vspace{-.2cm}
\subsection{Checkpoint Attribution System}

The previous works of Wang et al.~\cite{junichi_vocoders} and Pascu et al.~\cite{pascu_is24} found that large SSL models are better at discriminating between real and fake speech samples. Yet, it may be the case that "smarter" built models may not require as many parameters to incorporate similar information. Therefore we analyse one of the latest wav2vec-based models: \bert~\cite{seamless2023}.\footnote{\scriptsize{\url{https://huggingface.co/facebook/w2v-bert-2.0}}} The model is lighter (600m parameters) and showed increased performance over various speech classification tasks. 
As proxy, we select the best deepfake detection model from~\cite{pascu_is24}, i.e. \xls\footnote{\scriptsize{\url{https://huggingface.co/facebook/wav2vec2-xls-r-2b}}}. 
Following the approach of Pascu et al.~\cite{pascu_is24}, we extract audio features from the frozen SSL model and compute their temporal average. Consistent with our green AI objective, we adopt a simple k-Nearest Neighbors (kNN) model based on Euclidean distance. This model serves to investigate whether the SSL-derived features possess innate properties suitable for the task of audio deepfake model attribution.
We experimented with various values of $k$ and identified 
$k=21$ as the optimal trade-off between accuracy and computational complexity.
Due to space constraints we only report the macro F1-scores as objective measure, and refer the readers to our repository for more numeric results. 

\begin{table}[t!]
    \scriptsize
    \setlength{\tabcolsep}{3pt}
    \newcommand{\na}{{\color{gray}\textsc{n/a}}}
    \newcommand{\ii}[1]{{\scriptsize \color{gray} #1}}
    \newcommand{\lbl}[1]{{\scriptsize \color{gray} #1}}
    \newcommand{\key}[1]{{\footnotesize \texttt{#1}}}
    \newcommand{\dur}[2]{#1{\scriptsize ±#2}}
    \centering
    \caption{%
        An overview of the audio deepfake datasets used in this work.
        The Checkpoints column indicates the number of individual generative systems listed in the datasets' metadata.
    }
    \vspace{-.1cm}
    \label{tab:dataset}
    \begin{tabular}{llccrrr}
\toprule
Dataset                   &       & Langs.    & Checkpoints   & Spks   & Utts        & Duration\\
\midrule
\textbf{ASV19}&~\cite{asv19}               & en         & 17        & 78          & 63k         & \dur{3.1}{2.9} \\
\textbf{ASV21}&~\cite{asv21}               & en         & 97        & 14          & 49k         & \dur{2.9}{2.4}  \\
\textbf{ASV5}&~\cite{wang24_asvspoof}      & en         & 32        & 1159        & 410k        & \dur{9.5}{4.5} \\
\textbf{MLAAD v5}&~\cite{muller2024mlaad}  & 38          & 82        & \na          & 154k         & \dur{8.2}{9.0} \\
\textbf{TIMIT}&~\cite{timittts}              & en         & 15        & 9           & 20k         & \dur{3.1}{2.3} \\

\bottomrule
    \end{tabular}
    
    \vspace{-.5cm}
\end{table}

\begin{table*}[h!]
\caption{%
       Mean F1-scores $\uparrow$ across 3 random seeds for the kNN ($k=21$) checkpoint attribution model with \textbf{\xls} features. Results for all intermediate layers (columns) using different number of support samples to fit the kNN (rows). 
    }
    \vspace{-0.2cm}
    \centering
    \includegraphics[width=\linewidth]{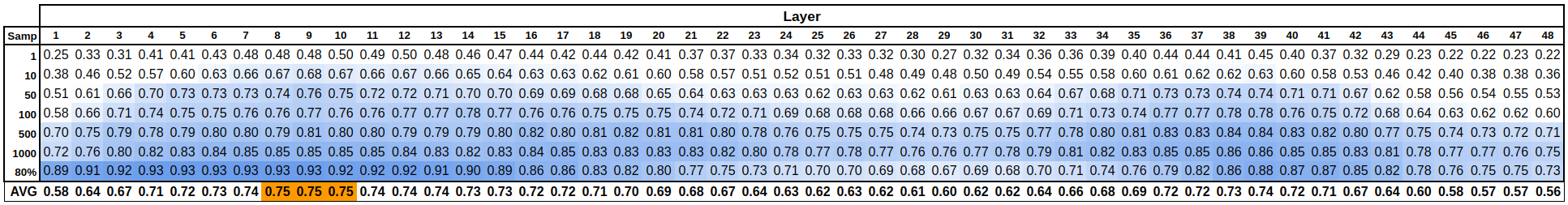} \\
    
    \label{fig:scores_xls}
    \vspace{-0.5cm}
    \end{table*}

\begin{table}
 \caption{%
      Mean F1-scores $\uparrow$ across 3 random seeds for the kNN ($k=21$) checkpoint attribution model with \textbf{\bert} features.  Results for all intermediate layers (columns) using different numbers of support samples in the kNN (rows). 
    }
    \centering
    \includegraphics[width=\linewidth]{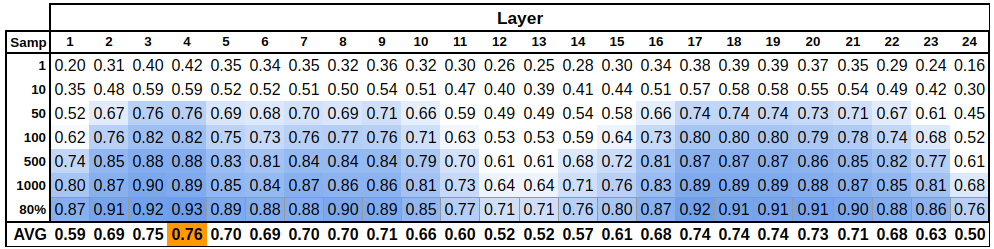}\\    
   
    \label{fig:scores_bert}
    \vspace{-.5cm}
\end{table}

\section{Evaluation and Analysis}

\subsection{SSL Model and Layer Selection}

The first evaluation step is to verify that the SSL models used for deepfake detection can also perform accurate model attribution. We compare the performance of two selected SSL models, as well as their intermediate representations (layer-wise)~\cite{zhang2024audio,martindonas24_interspeech}. And test their ability to perform the task with different numbers of support samples.\footnote{To emphasize that there is no training involved in instance-based models, we use the \emph{support set} notion instead of \emph{training set }.} Tables~\ref{fig:scores_xls} and \ref{fig:scores_bert} depict the F1-scores obtained for each intermediate layer of the two models (columns) when selecting different numbers of samples from each checkpoint to fit the kNN (rows). The number of neighbours is fixed to $k=21$, and all checkpoints are seen in the support set, i.e. in-domain attribution. There are a total of $243$ classes (checkpoints) to predict. 
We observe very high F1-scores even from the early layers of both models, with as few as 50 samples per model checkpoint: \bert model at layer 4 achieves an F1-score of $0.76$, and the \xls model at layer 9 attains the same F1-score. When 80\% of the checkpoint samples are used for fitting the kNN, the F1-scores increase to $0.91$ for both models as early as the second layer. Compared to the results of Müller et al.~\cite{muller22b_interspeech} ($97.10\%$ accuracy) and Phykan et al.~\cite{phukan2024investigatingprosodicsignaturesspeech} ($98.91\%$ accuracy), our kNN-based approach achieves perfect accuracy ($100\%$) and F1-score ($1.0$) for ASV19 over three 80:20 random splits.

The final rows in Tables~\ref{fig:scores_xls} and \ref{fig:scores_bert} present the average F1-score across different sample sizes for each layer. Notably, the \bert model outperforms the \xls model, with layer 4 yielding the highest average F1-score ($0.76$). In line with our green AI focus, we select the \bert model for all subsequent experiments, and extract the features from the best layer. As a result, we use 121m parameters ($\approx6\%$ of the complete \xls model) and attain an average 14 msec inference time for 10 second-long audio chunks at 1.13 GB VRAM usage on a Tesla V100 GPU. 

\vspace{-.1cm}

\subsection{Checkpoint Attribution Analysis}
\label{sec:base}

\begin{figure}[t!]
    \centering
    \includegraphics[width=1\columnwidth]{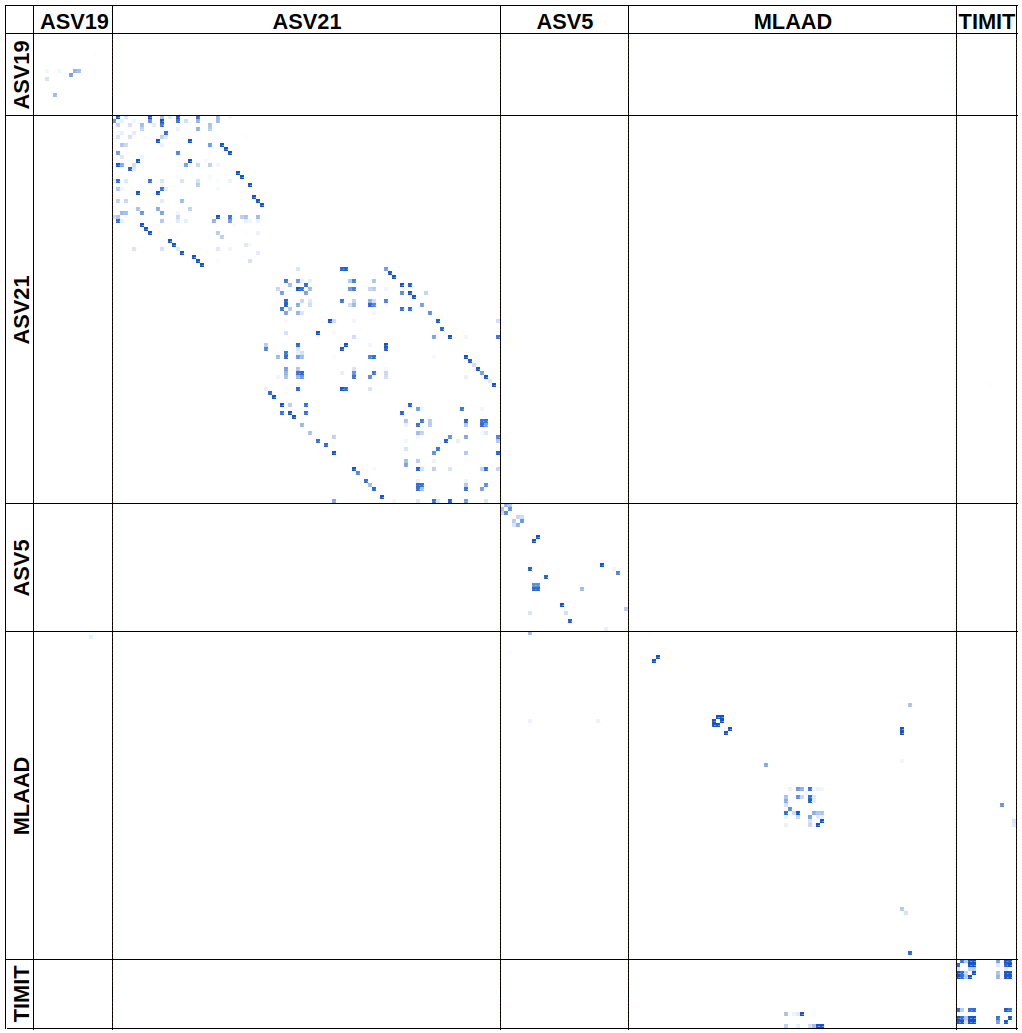}\\
   
    \caption{%
       Percent of samples (colour shades) in the 21-neighbour vicinity of the \bert features-based kNN which do not pertain to the 
       target class. Main diagonal values (same checkpoint) and individual checkpoint names are omitted for readability. 
    }
    \label{fig:neighbors}
    \vspace{-.6cm}
\end{figure}

A generative model may be part of multiple deepfake detection datasets.
For example, both MLAAD and TIMIT contain samples generated by VITS.
However, the VITS model has multiple checkpoints and it is unclear which ones were used by the two datasets.
Since the deepfake datasets do not provide this information, we treat each
model id listed in the datasets as an independent source of synthetic audio.
%
To gain a deeper understanding of the large and diverse dataset we are using, we conduct an exploratory analysis of its samples.
We go back to our kNN approach and examine the 21 closest neighbours of each sample. We count how many of these neighbours originate from a different checkpoint.
Figure~\ref{fig:neighbors} illustrates these results. The darker shades indicate a higher number of non-target samples. Due to the large number of checkpoints, we only list the datasets in the figure (the full matrix is available in the code repository). Additionally, we note that the matrix is not necessarily symmetrical, as the closest samples in the immediate vicinity may not be the same. For clarity, we omit the main diagonal values (same checkpoint), as they are not relevant to this analysis.

The first observation is that there is very little confusion between the samples pertaining to the different datasets. The single exception is for MLAAD and TIMIT. Upon further investigation, we find that these confusions primarily occur for the VITS and Speedy Speech checkpoints. In the case of TIMIT Speedy Speech, the confusion is notably high, with 46\% of the neighbours coming from the MLAAD checkpoint. This suggests that the dataset developers may have used the same checkpoint to generate the samples. Additionally, for TIMIT and MLAAD, more non-target neighbours appear for checkpoints using LJSpeech as the synthesis identity. The LJSpeech checkpoints also contribute to intra-dataset confusions. For TIMIT, the only checkpoints that do not exhibit non-target samples in the nearby vicinity are those from the multi-speaker models. This observation may indicate that the \bert features are encoding the speaker information to a large extent.

For ASV21, the separate regions in the plot are a result of the different sources of the checkpoints, i.e. VCC2018 and VCC2020 audio samples. We still observe a considerable number of non-target neighbours, as well as a certain symmetry among them. Both challenges involved two main tasks (parallel and non-parallel or cross-lingual voice conversion), and most teams submitted the same conversion system for both tasks. Additionally, many submissions were very similar in terms of the architecture and/or vocoder. The fact that the \bert model was able to capture these similarities is quite surprising. However, it may be attribute that these non-target neighbours pertain to the same speaker identities used for voice conversion.

In ASV5 we have variants of the same TTS architecture: GlowTTS (checkpoints A01, A02, A03); GradTTS  (A04, A05, A06); and ToucanTTS (A09, A10, A21, A22).\footnote{Samples altered by adversarial attacks are not considered.} Based on informal listening tests, these checkpoints generate different speaker identities. Yet, the \bert features capture neighbouring samples within the same speech generation architecture. This suggests that we may also perform architecture attribution using kNN, which is another promising indication of the model's capabilities.


With the confusions observed for the LJSpeech samples, we wonder if the entire checkpoint attribution is not in fact a speaker attribution system--given that most checkpoints use different speaker identities.
Therefore, we first do a contracted checkpoint attribution over the systems trained only on LJSpeech---a single speaker dataset. A total of 20 checkpoints were selected (see our code repository). The same kNN with $k=21$ is fitted. We split the data into 80:20 ratio and obtain an $0.74$ F1-score for the attribution. If this were a mere speaker attribution, we would have seen a much lower performance. 
We also examine the multi-speaker models, which, from a checkpoint perspective represent the same source. We select the multispeaker models from TIMIT and MLAAD, and separate them based on speaker and language (a different label for each identity). We obtain an $0.72$ F1-score for this attribution. This is in contrast to the $0.99$ F1-score we obtain when not splitting the multi-speaker checkpoints. 
So it appears that although the speaker attributes are captured by the representations, other architecture-indicative signal characteristics are inherent to the \bert features.

\vspace{-.2cm}
\subsection{Unseen Checkpoint Detection}

Speech generation is an ever-evolving field, with new systems being introduced regularly. Framing the task of deepfake source tracing as a closed classification problem is insufficient for practical, end-user applications. To address this scenario, we investigate how our method can also be applied to detect unseen or out-of-domain (OOD) checkpoints~\cite{lu2023detecting}. Here, we define out-of-domain data as checkpoints for which several samples are available, but no labels have yet been assigned to them. 

Vaze et al.~\cite{vaze2022openset} showed that out-of-domain classification can be efficiently performed when a robust closed-set classifier is available. We already have such a classifier, and we simulate out-of-distribution detection by randomly setting aside four checkpoints from each of our five datasets (i.e., 20 checkpoints out of a total of 243). Half of these checkpoints are used for validation, and half for evaluation, with no overlap between the sets. The remaining in-domain class samples are then split into support, validation, and test subsets with an 80:10:10 ratio.
We fit the kNN model over the support set of in-domain samples. 
 Next, we compute the average Euclidean distance of all validation data points to their 
$k$ closest neighbours and establish an OOD detection threshold based on the Equal Error Rate (EER) of this subset. The same average distance is then computed for the test samples. The final OOD decision is made relative to the threshold derived from the validation set.

The mean and standard deviation of the F1-score across three different OOD partitions are presented in Table~\ref{tbl:knn_ood}. The results are shown in terms of the number of neighbours (rows), with individual dataset results, as well as results for the entire test set (columns). For the OOD evaluation on individual datasets, we compare the OOD samples from each dataset against all in-domain samples from all datasets, meaning that a single OOD detector is trained, and its predictions are processed accordingly. We observe a relatively high average F1-score of $0.84$ ($k=21$) across all datasets, with some variations among them. Best results are for $k=1$, but in practice this may not be a feasible option due to improper labelling. 
The performance for individual datasets appears strongly correlated with the results from Section~\ref{sec:base}: a higher number of non-target neighbours for a dataset's checkpoints translates into a worse OOD detection performance.

For the OOD checkpoints, the previous section showed that as little as 10 random labelled samples in the support set would yield an F1-score of $0.59$ for in-domain attribution.

\begin{table}[t!]
\centering
\footnotesize
\newcommand{\ii}[1]{{\scriptsize \color{gray} #1}}
\setlength{\tabcolsep}{1.5pt}
\caption{Mean and standard deviation of the F1-scores $\uparrow$ reported across 3 random seeds for the kNN-based \textbf{OOD detection}. We use different numbers of neighbours (rows) and report results for the complete (All) and individual subsets (columnns). }
\label{tbl:knn_ood}
\vspace{-.1cm}
\begin{tabular}{r|ccccc|c}
    \toprule
    \textbf{k}  & \textbf{ASV19} &\textbf{ASV21}&\textbf{ASV5}&\textbf{MLAAD}&\textbf{TIMIT} & \textbf{All} \\ \midrule
     $1$             &   0.87±0.02     & 0.58±0.12    &  \textbf{0.86±0.06 } & \textbf{0.78±0.11}    & \textbf{0.70±0.07} & \textbf{0.88±0.03}  \\
     $5$             &   \textbf{0.88±0.04}     & 0.58±0.10     & 0.84±0.06  & 0.74±0.14    & 0.69±0.07 & 0.86±0.04  \\
     $21$            &   0.86±0.03     & \textbf{0.58±0.08}     & 0.81±0.07  & 0.71±0.15    & 0.66±0.06 & 0.84±0.04\\ \bottomrule

\end{tabular}

\vspace{-.1cm}
\end{table}



    

\vspace{-.2cm}
\subsection{Beyond checkpoint attribution}

\begin{table}[t!]
\centering
\footnotesize
\newcommand{\na}{{\color{gray}\textsc{n/a}}}
\newcommand{\ii}[1]{{\scriptsize \color{gray} #1}}
\setlength{\tabcolsep}{1.5pt}
\caption{F1-scores $\uparrow$ for the kNN-based \textbf{acoustic model and vocoder classification} with different datasets and data splits. The splits in MLAAD3 are those reported in~\cite{klein2024source}. \emph{AM} stands for acoustic model, and \emph{Voc} stands for vocoder attribution. The \emph{Split} column refers to the use of the same checkpoints for fitting and evaluation (ID) or setting aside one (OOD-1) or half of the checkpoints (OOD-H) from the support set. \emph{MLD3} = MLAAD v3; \emph{MLD5} = MLAAD v5;  and \emph{TIM} = TIMIT. 
}
\label{tbl:hemlata}
\vspace{-.2cm}

\begin{tabular}{cl|l|c|cc}
    \toprule
    &\textbf{Support set} & \textbf{Test data} & \textbf{Split} & \textbf{AM} & \textbf{Voc} \\ \midrule
    \ii{0} & \multicolumn{3}{l}{MLD3 baseline~\cite{klein2024source}} &   0.82                  &  0.93          \\  \midrule
    \ii{1} & MLD3 train             &  MLD3 test            & ID            &   0.89                  &  0.90         \\ 
    \ii{2} & MLD3 train+dev     &  MLD3 test        & ID            &    \textbf{0.96}                &  \textbf{0.96}  \\ \midrule
    \ii{3} & ASV5+MLD5+TIM    &  ASV5+MLD5+TIM  & ID            &    0.99±0.01           &  \na            \\ 
    
   \ii{4} & MLD5               &  ASV5+TIM         & OOD           &    0.30±0.00            &  \na            \\ 
   \ii{5} & ASV5+MLD5+TIM    &  ASV5+MLD5+TIM  & OOD-1           &    0.55±0.19            &  \na            \\ 
   \ii{6} & ASV5+MLD5+TIM    &  ASV5+MLD5+TIM  & OOD-H           &    0.48±0.10           &  \na            \\
   \bottomrule								                                     
\end{tabular}

\vspace{-.5cm}
\end{table}

Previous work has also explored the provision of more explainable decisions or attributes for speech generation checkpoints~\cite{klein2024source,chhibber2024}. We follow the tasks outlined in~\cite{klein2024source} and use their evaluation protocol for MLAAD. It is important to note that Klein et al.~\cite{klein2024source} used version 3 of MLAAD, and for a direct comparison, we revert to that version in this experiment. 
Two files from the test set are discarded--they are very long ($>$90 seconds) and crashed the inference, and we do not use the genuine files for attribution purposes. The same front-end and kNN classifier ($k=21$) were employed. The results are presented in Table~\ref{tbl:hemlata}, rows $1$ and $2$, and clearly surpass the baseline (row $0$). 

Based on these results, it might be tempting to assume that identifying the acoustic model and vocoder is a trivial task. However, the evaluation in~\cite{klein2024source} is limited to intra-dataset in-domain scenarios. As indicated by our results in Section~\ref{sec:base} this is essentially a larger grouping of the intra-dataset samples. To further assess the robustness of our approach, we proceed to test \textit{cross-dataset} and \textit{out-of-domain} acoustic model attribution. Vocoder attribution is less reliable due to the information not being clearly marked in the datasets' metadata.
We select the following architectures common to ASV5, MLAAD v5, and TIMIT: FastPitch, GlowTTS, Tacotron, Tacotron2, VITS, and xTTS. First, we compute the F1-score using a random 80:20 train-test split ratio of the samples. The results are shown in Table~\ref{tbl:hemlata} row $3$. While the performance is notably high, this is the same in-domain attribution, and does not capture the challenge of cross-dataset or out-of-domain attribution.

Next, for a true out-of-domain acoustic model attribution, we use the checkpoints from MLAAD v5 to build the kNN support set, and test on TIMIT and ASV5. The results, shown in row $4$ of Table~\ref{tbl:hemlata} indicate a significant drop in performance (F1-score of $0.30$). The majority of samples are classified as xTTS, which is the dominant class in the support set (33k samples). To address the class imbalance, we experimented with undersampling approaches and Condensed kNN~\cite{1054155}, but the results remain similar. 
So it may be that MLAAD is not sufficiently diverse to generalise to the other datasets. We lastly employ a leave-N-out strategy, grouping all the samples from the datasets into six acoustic models, and randomly selecting one or more checkpoints from each architecture for testing. The results show that performance remains low, with an F1-score of $0.55$ when a single checkpoint is set aside for testing (row 5, OOD-1) and $0.48$ when half of the checkpoints are set aside (row 6, OOD-H). It is also important to mark the wide variance in the last two rows, which means that different unseen checkpoints pose different issues to the attribution system.

As a result, although checkpoint attribution appears to be a rather simple task for our training-free \bert features, unravelling the underlying architecture or vocoder for the model may require alternative classification strategies, as well as most-likely more fine-grained audio datasets.  





\vspace{-.2cm}
\section{Conclusions}

In this work we showed that using a powerful pretrained SSL model alleviates the need for any additional, complex back-end modules to accurately predict the source of a generated speech sample. 
We also showed that early layers of the SSL models are sufficiently informative for kNN-based checkpoint attribution ($0.93$~macro F1-score). The same setup can be successfully applied to detect samples from out-of-domain checkpoints, unseen in the kNN support set ($0.84$~F1-score). As a result, our framework is lightweight and training-free, in line with green AI endeavours. 

We argue that our results support the exploitation of pretrained models as more than just naive feature extractors. They should rather be considered as a fundamental source of information that needs to be further understood, perhaps in a more traditional manner (e.g. feature analysis).
As future work, we would like to understand how minor checkpoint changes may impact this very accurate attribution. For example, would a few-shot adaptation process for a TTS model change its attribution? How about the outputs from earlier epochs of the training?

\clearpage
\textbf{Acknowledgement.} This work was co-funded by EU Horizon project AI4TRUST (No. 101070190), and by the Romanian Ministry of Research, Innovation and Digitization project DLT-AI SECSPP
(ID: PN-IV-P6-6.3-SOL-2024-2-0312).

\bibliographystyle{IEEEtran}
\bibliography{mybib}

\end{document}